\documentclass[12pt]{article}
\usepackage{epsfig}
\usepackage{amsmath}
\usepackage{bm}
\usepackage{graphicx}
\begin{document}

\begin{center}
{\bf FIRST-PRINCIPLES CALCULATIONS OF $LiNbO_3$ OPTICAL PROPERTIES:
FROM FAR-INFRARED TO ULTRAVIOLET}

\vspace{5mm}

{VLADIMIR SALEEV}

{Department of Physics, Samara National Research University,\\
 Moskovskoe Shosse, 34,
Samara, 443086, Russia\\
saleev@samsu.ru} \vspace{5mm}

 {ALEXANDRA SHIPILOVA}

{Department of Physics, Samara National Research University,\\
 Moskovskoe Shosse, 34,
Samara, 443086, Russia\\
alexshipilova@samsu.ru}
\end{center}

\begin{abstract}
We perform first-principles calculations of optical properties for
ferroelectric phase of $LiNbO_3$ crystal using density functional
theory for wide range of wavelengths, from far-infrared to
ultraviolet. We study frequency dependence of complex dielectric
tensor and related quantities, such as refractive and reflection
indices, absorption coefficients, etc. Our calculation incorporate
advantages of numerical approaches based on atomic-orbital
all-electron Gaussian-type basis sets, as it is realized in
CRYSTAL14 program. We compared predictions obtained in
general-gradient approach with PBESOL exchange-correlation
functional and in hybrid approach with PBESOL0 functional, and we
have found that hybrid PBESOL0 functional is more applicable to
describe the wide set of the experimental data.
\end{abstract}


\section{Introduction}

{Lithium Niobate (LN) crystallizes into two different phases,
ferroelectric and paraelectric, depending on the temperature}. The
ground state of LN (space group $R3c$) undergoes the phase
transition at the temperature near 1480 K to the high-symmetry
paraelectric phase (space group $R\bar{3}c$). The ground-state phase
has very large dielectric, piezoelectric, pyroelectric, non-linear
optic and electro-optic responses. That is why it is well-known as
excellent material for different branches of photonics. The results
of experimental study of LN main physical properties are collected
in Ref. \cite{LN1985}, and the more recent references {to} data can
be found in Springer materials database \cite{LNwww}. {The number of
LN parameters} relevant for electro- and elasto-optics from
different measurements have been recalculated and determined in Ref.
\cite{LN2002}. Complete sets of elastic constants and photoelastic
coefficients of LN crystal are presented in Ref. \cite{LN2009}.
Infrared reflection spectra and Raman spectra of LN were studied
many years ago for the first time in Refs. \cite{LN65,LN68}. Up to
now, Raman frequencies of LN were experimentally studied many times,
see Refs. \cite{LN62,LN63,LN64,LN66,LN67,LN69,LN70}. Indices of
refraction and absorption in far-infrared range between 100 GHz and
3 THz (3 cm$^{-1}$ - 100 cm$^{-1}$) were investigated in Ref.
\cite{LNTHz}. Light absorption in the visible wavelength range was
measured in Ref. \cite{LNvisible}.

Theoretical first-principles study of LN optical properties in the
framework of Density Functional Theory (DFT) \cite{DFT1,DFT2} are
based on linear response theory. At first time, {\it ab initio}
calculations of LN crystal phonon spectrum, infrared (IR) and Raman
frequencies, were performed in Ref. \cite{PRB2000}. Electro-optic
effects in LN crystal were calculated in Ref. \cite{PRL2004}, where
authors explicitly took into account the electronic, ionic and
piezoelectric contributions. The precise calculations of the
electron band structure and the dielectric function of ferro- and
paraelectric LN with inclusion of quasiparticle and electron-hole
effects in the range of visible and ultraviolet (UV) wavelengthes
were studied in Refs. \cite{PRB2008,CMS2013}. The influence of the
nonlocal Hartree-Fock exchange on the band gap value of LN in the
framework of hybrid DFT was investigated in Ref. \cite{PRB2014}.
Recently, theoretical study of electronic and optical properties of
LN under high pressure was performed using first-principles methods
in the Ref. \cite{China2015}. {The influence of electronic many-body
interactions, spin-orbit coupling, and thermal lattice vibrations on
the electronic structure of LN were calculated by combining hybrid
density functional theory with the $QSGW_0$ scheme in the work
\cite{LNGW2016}.}

In this paper we perform first-principles calculations for complex
components of dielectric tensor of ferroelectric LN crystal and
relevant optical parameters exploring CRYSTAL14 programm package
\cite{cry14}. We study frequency dependence of considered optical
properties in the wide region of wavelengths, starting from
far-infrared (including static limit) region to visible and soft UV
waves. We consistently merged ionic and electronic contributions to
describe intermediate region where both ones are important. The
paper is organized as follows. We describe relevant physical models
in the Sec. 2. In the Sec. 3 we present computational methods and
program input. In the Sec. 4 we discuss obtained results and perform
comparison with experimental data and previous calculations.
Finally, we summarize our conclusions in the Sec. 5.

\section{Physical model}

The main optical properties of crystal, such as refraction and
absorption indices, reflectivity, Raman spectrum, are connected with
permittivity of crystal, which is characterized by means of a
complex dielectric tensor $\boldsymbol{\varepsilon}$. Such a way, if
we know real $(\boldsymbol\varepsilon_1)$ and imaginary
$(\boldsymbol\varepsilon_2)$ parts of dielectric tensor,
$\boldsymbol\varepsilon=\boldsymbol\varepsilon_1+i
\boldsymbol\varepsilon_2$, we can calculate real and imaginary parts
of complex refraction index, $\sqrt{\varepsilon}=n^*=n+i k$, where
$$n=\left[ \frac{(\varepsilon_1+\varepsilon_2)^{1/2}+\varepsilon_1}{2}\right]^{1/2} \mbox{ and }
k=\left[
\frac{(\varepsilon_1+\varepsilon_2)^{1/2}-\varepsilon_1}{2}\right]^{1/2}.$$
Classical reflectivity is defined as
$$R=\frac{(n-1)^2+k^2}{(n+1)^2+k^2},$$
and the absorption coefficient is written $$\alpha=\frac{4\pi
k}{\lambda}.$$

 Wavelength and
wavenumber (or frequency) are presented via photon energy as follows
$$\lambda(\mu m)  =\frac{1.2398}{E (eV)}, \quad  \tilde{\nu}(cm^{-1})=\frac{1}{\lambda}=\frac{E(eV)}{1.2398 \cdot 10^{-4}}$$

Complex dielectric tensor $\varepsilon_{ii}({\tilde{\nu}})$ written
in diagonal form can be presented for each inequivalent polarization
direction as a sum of electronic ("high-frequency") and ionic
("low-frequency") contributions :
\begin{equation}
\varepsilon_{ii}(\tilde{\nu})=\varepsilon_{el,ii}(\tilde{\nu})+\varepsilon_{ion,ii}(\tilde{\nu}).\label{equ1}
\end{equation}
Ionic contribution is written in semi-classical oscillator model
which applied for infrared region ($\lambda >10^4$ nm or
$\tilde{\nu}<10^3$ cm$^{-1}$):
\begin{equation}
\varepsilon_{ion,ii}(\tilde{\nu})=\sum_p\frac{f_{p,ii}\tilde{\nu}^2_p}{\tilde{\nu}_p^2-\tilde{\nu}^2-i\tilde{\nu}
\gamma_p}
\end{equation}
where $ii$ indicates the polarization direction, $\tilde{\nu}_p$,
$f_p$ and $\gamma_p$ are the transverse optical frequency,
oscillator strength and damping factor for the $p$-th vibration
mode, respectively. The set of harmonic phonon frequencies
$\tilde{\nu}_p$ in the the $\Gamma$ point can be obtained from the
diagonalization of the Hessian matrix of the second derivatives with
respect to atomic displacements \cite{cry14_2}:
\begin{equation}
H^{\Gamma}_{ai,bj}=\frac{1}{\sqrt{M_a M_b}}\left( \frac{\partial^2
E}{\partial u_{ai}\partial u_{bj}} \right)\label{equH}
\end{equation}
where $M_{a,b}$ are the atomic masses,  $u_{ai}$ and $u_{bj}$ are
displacements of atoms $a$ and $b$ in the reference cell  along the
$i$-th and $j$-th Cartesian directions, respectively. $E$ is the
total crystal energy, as it is calculated in DFT at the fixed ionic
positions. In standard way, the crystal energy in (\ref{equH}) is
calculated at the zero external electric field and it is supposed
that oscillator strengths don't depend on wavelength and intensity
of scattering light as it should be in the linear approximation. We
will follow this approximation taking into account that experimental
data correspond to finite values of wavelength and intensity of
light. {We mention that} the exact calculation of such non-linear
response, which may be not small for ferroelectric crystals, is not
a matter of standard approach implemented in CRYSTAL14 package.

In the static limit ($\tilde{\nu}\to 0, \lambda \to \infty$), which
is realized accordingly our calculations at the $\lambda \geq 5
\cdot 10^5$ nm ($\tilde{\nu} \leq 20$ cm$^{-1}$), the components of
the dielectric tensor $\varepsilon$ are real constants,
$$\varepsilon_{ii}(0)=\varepsilon_{el,ii}(0)+ \sum_p f_{p,ii}.$$
In the range of $5\cdot 10^3 \leq \lambda \leq 5\cdot 10^5$ nm ($20
\leq \tilde{\nu} \leq 2 \cdot 10^3$ cm$^{-1}$) components of
dielectric tensor are calculated {by the formula} (\ref{equ1}) with
$\varepsilon_{el,ii}(\tilde{\nu})\approx \varepsilon_{el,ii}(0)$.

 The high-frequency region includes visible and ultraviolet wavelengths up to frequencies
 which correspond to an electronic excitation of separate atoms,
  $\lambda \leq 10^4 \mbox{ nm}$  ($\tilde{\nu}\geq 10^3 \mbox{cm}^{-1}$).
  Here, the ionic contribution vanishes and dielectric tensor
is computed in a quasi-free electron approximation via coupled
perturbed Hartree-Fock/Kone-Sham (CPHF/CPKS) method \cite{cry14_3}.
It is a perturbative, self-consistent method that focuses on the
description of the relaxation of the crystalline orbitals under the
effect of an external electric field. The perturbed wave function is
then used to calculate the dielectric properties as energy
derivatives.

{\it Ab initio} calculations of photoelastic (elasto-optic)
constants are developed and implemented in the CRYSTAL14 program
\cite{Erba2013}. These constants are the elements of the fourth rank
photoelastic (Pockels) tensor and are defined as:
\begin{equation}
p_{uv}=\frac{\partial\triangle\varepsilon_{u}^{-1}}{\partial\eta_{v}},\label{pockel}
\end{equation}
where $\triangle\varepsilon_{u}^{-1}$ is the difference of the
inverse dielectric tensor between strained and unstrained
configurations, $\eta$ is the rank-2 symmetric tensor of pure strain
and Voigt's notation is used according to which $u,v=1,...,6\quad (1
= xx, 2 = yy, 3 = zz, 4 = yz, 5 = xz, 6 = xy)$. The dielectric
tensor of the equilibrium structure and of each strained
configuration is computed via a CPHF/KS scheme in the two regimes.
The first one is an infinite wavelength approximation (static limit)
and the second one takes into consideration finite wavelength of the
external electric field.

To demonstrate self-consistency, completeness and {good agreement}
with experimental data of our DFT-based calculations, we also
calculate elastic \cite{elast} and piezoelectric \cite{piezot}
tensors for LN crystal and compare results with {existing} data.
Elastic constants are calculated as derivatives of total energy per
{unit} cell with volume $V$,
\begin{equation}
C_{uv}=\frac{1}{V}\left(\frac{\partial^2 E}{\partial\eta_u
\partial\eta_v}\right)_{\eta_u=\eta_v=0}.
\end{equation}

Piezoelectric tensors $\bf{e}$ describe the polarization $\bf P$
induced by strain $\eta$. The Cartesian components of the
polarization $P_i$ can then be expressed in terms of the strain
tensor components: $$ P_i=\sum_v e_{iv}\eta_v,$$ so that at the
constant field $\bf\cal E$ as follows
\begin{equation}
e_{iv}=\left( \frac{\partial P_i}{\partial \eta_v} \right)_{\cal E}.
\end{equation}
In CRYSTAL14, the polarization can be computed either via localized
Wannier functions or via the Berry phase (BP) approach. The latter
scheme is used in the present automated implementation according to
which direct piezoelectric constants can be written as follows in
terms of numerical first derivatives of the BP $\phi_l$ with respect
to the strain:
$$e_{iv}=\frac{|e|}{2\pi V}\sum_l a_{li}\frac{\partial \phi_l}{\partial\eta_v},$$
where $a_{li}$ is the i-th Cartesian component of the l-th direct
lattice basis vector ${\bf a}_l$.

\section{Computation methods and program input}

During the self-consistent energy calculations we use DFT
functionals of different types: generalized gradient approximation
(GGA), such as PBESOL \cite{pbesol}, and hybrid, such as PBESOL0
\cite{pbesol} (with 25 \% of Hartree-Fock exchange mixing).

All-electron atom-centered Gaussian-type function basis sets are
adopted for Lithium and for Oxygen, {\tt Li-5-11(1d)G} and {\tt
O-8-411(1d)G}, correspondingly. Effective core pseudo-potential
approach is used for Niobium atom with  Hay and Wadt small core
potential (HAYWSC) and four-valence-electron basis set is {\tt
Nb-SC-HAYWSC-31(31d)G}. The used basis sets have been optimized for
LN crystal in Ref. \cite{baranek}.

The accuracy of calculating the energies of Coulomb and Hartree-Fock
exchanges is controlled in the CRYSTAL14 program by a set of {\tt
TOLINTEG} parameters, which were chosen as {\tt \{8, 8, 8, 8, 30\}}.
The convergence threshold on energy for the self-consistent-field
(SCF) calculations is set to $10^{-7}$ Hartree for structural
optimization and $10^{-8}$ Hartree for vibration frequency
calculations.  The number of basis vectors in the irreducible
Brillouin zone is given by the shrink parameter IS = 8 for
structural optimization and IS=16 for vibration frequency and
CPHF/KS calculations. The relaxation of cell parameters and atomic
positions to equilibrium values was carried out until the lattice
stress became less than 0.02 GPa.

Ionic contribution to the dielectric tensor is calculated as
function of frequency together with infrared spectra by means of the
following scripts: {\tt FREQCALC, INTENS, DIELTENS, IRSPEC, DIELFUN,
REFRIND}. Electronic high-frequency component of dielectric tensor
is calculated by script {\tt CPKS}, which includes DFT in both its
local and generalized-gradient PBESOL approximations, and to hybrid
functional PBESOL0.

To perform elastic, piezoelectric or photoelastic calculations we
use {\tt ELASTCON, ELAPIEZO} and {\tt PHOTOELA} scripts with {\tt
PREOPTGEOM} option. The last one performs additional test and
precise geometry optimization before the step of properties
calculation.

\section{Results}

\subsection{Crystal structure and relaxation}

The lattice parameters of relaxed structures are very close to the
experimental values with a few percent accuracy. As it is estimated,
the calculation with PBESOL GGA exchange-correlation functional
gives very close to experimental value for density, taking in mind
the temperature dependence of lattice constants and density. Results
obtained with hybrid functional PBESOL0 are also not so far from the
experimental value. In the Table \ref{tab1}, the atomic positions of
irreducible  atoms, obtained with three different functionals, are
also collected and compared with experimental data from Ref.
\cite{latice}. {The predicted fractional coordinates correspond to
experimental ones within the approximate accuracy of 1 \%}. There is
only one exception for the $y-$coordinate of oxygen atom which has a
slightly larger difference than others.

\begin{table}[ph]   
\caption{Crystallographic information: lattice constants $a$ and
$c$; density; positions of irreducible atoms.}
\begin{center}
{\begin{tabular}{@{}lcccccc@{}} \hline
\\[-1.8ex]
EC functional & $a$, {\AA} &
$c$, {\AA} & $\rho$, g/cm$^3$ & Nb & Li & O \\[0.8ex]
\hline \\[-1.8ex]
PBESOL & 5.118 & 13.958 & 4.653 & \{0,0,0\} & \{1/3, 2/3, 0.9536\} & \{0.3187, 0.0237, 0.8962\}\\
PBESOL0 &5.084 & 13.842 & 4.756 &\{0,0,0\} & \{1/3, 2/3, 0.9536\} & \{0.3190, 0.0247, 0.8960\}\\
Experiment \cite{latice}& 5.151 & 13.876 & 4.648 & \{0,0,0\} & \{1/3, 2/3, 0.9513\} & \{0.3239, 0.0383, 0.8983\}\\[0.8ex]
\hline \\[-1.8ex]
\end{tabular}}
\end{center}
\label{tab1}
\end{table}

\subsection{Elastic constants and phonon dispersion}

\begin{figure}[h]
\centerline{\psfig{file=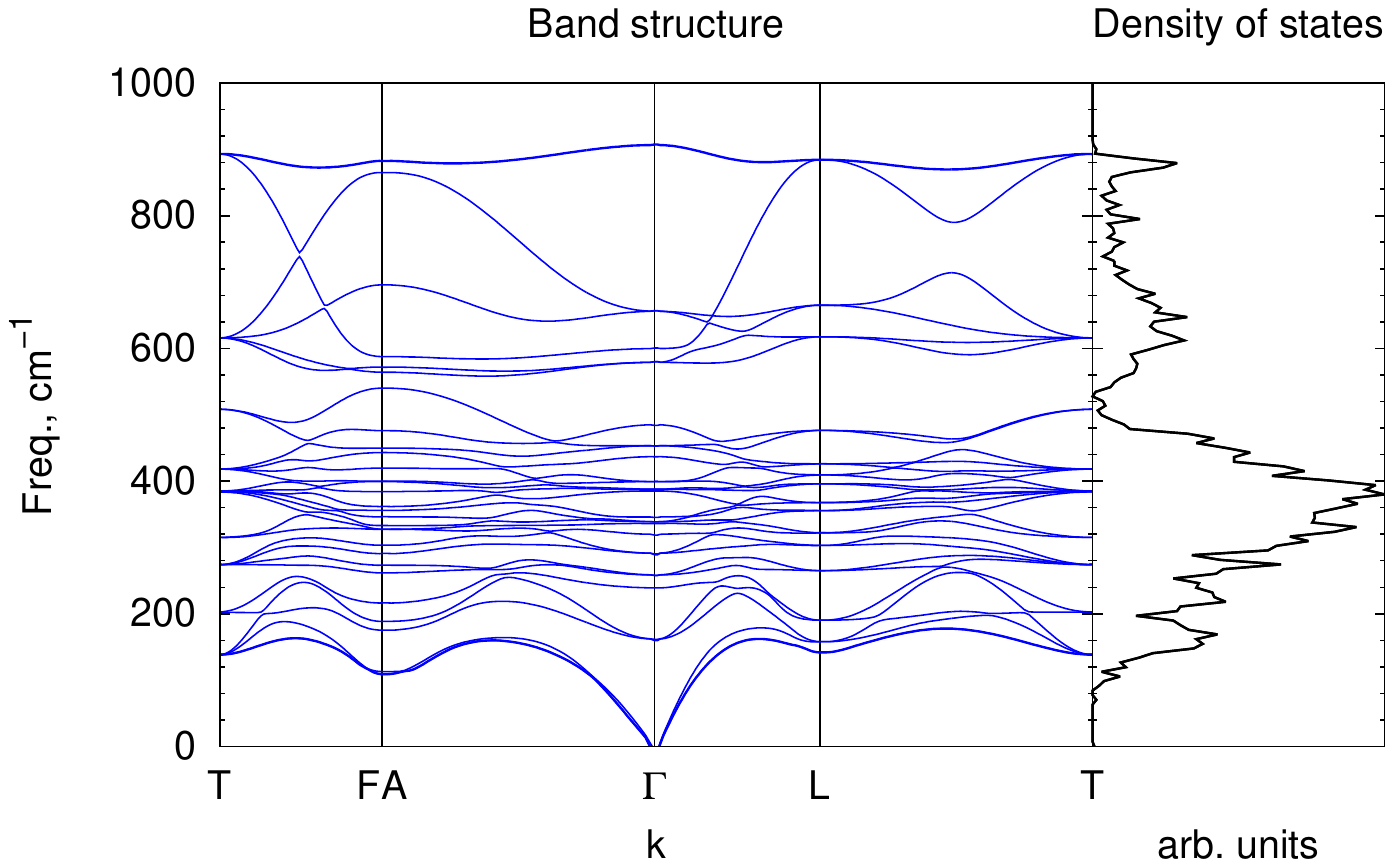,width=6.00in}}
\vspace*{8pt} \caption{Phonon band structure and DOS for LN crystal
obtained with PBESOL0 functional.} \label{fig:phonon}
\end{figure}

The general tests of mechanical stability of relaxed theoretical
structure {are based on the requirement of the positivity of
eigenvalues of their elastic constants $C_{uv}$ matrix}. {Also, the
calculation and plotting of phonon dispersion curves is to be very
informative}. The sets of elastic constants obtained with PBESOL and
PBESOL0 functionals  are collected in the Table \ref{tab2} as well
as the experimental data. In the Fig. \ref{fig:phonon}, phonon band
structure and density of states (DOS) along the high-symmetric
crystallographical directions are shown in case of calculation with
PBESOL0 functional. We consider these calculations corresponding to
the case of zero external electric field.

\begin{table}[h!]   
\caption{Elastic constants $C_{uv}$ in GPa at constant field ${{\cal
E}}$.} \begin{center} {\begin{tabular}{@{}lccccccc@{}} \hline
\\[-1.8ex]
EC functional & $C_{11}$ &
$C_{12}$ & $C_{13}$ & $C_{14}$  & $C_{33}$ & $C_{44}$ & $C_{66}$ \\[0.8ex]
\hline \\[-1.8ex]
PBESOL & 194&64 &67 & 16&221 &45 &63\\
PBESOL0 & 213&66 &70 & 18&242 &51 &71\\
Experiment \cite{cab} &203 & 53&75 &9 & 245& 60&75\\
Experiment \cite{LN2009} &200 & 56&70 &8 & 240& 60&72\\[0.8ex]
\hline \\[-1.8ex]
\end{tabular}}
\end{center}
\label{tab2}
\end{table}

\subsection{Piezoelectric constants}

 Here we present the results of calculations for piezoelectric
 constants $e_{iv}$ for hexagonal case, in $C/m^2$, and compare it
 with the most recent experimental data \cite{piezo}. The more
 complete sets of experimental data can be found in
 Ref.~\cite{LN2009}. Mostly, our predictions coincide to the data with relative errors of
about 5-10 \%.

\begin{table}[h!]   
\caption{Piezoelectric constants $e_{iv}$ in $C/m^2$.}
\begin{center}{\begin{tabular}{@{}lcccc@{}} \hline
\\[-1.8ex]
EC functional & $\quad e_{15}$   &
$\quad e_{22}$   & $\quad e_{31}$   & $\quad e_{33}$ \\[0.8ex]
\hline \\[-1.8ex]
PBESOL & 3.62 & 2.23 & 0.17 & 1.24\\
PBESOL0 & 3.71 &2.20 & 0.20 & 1.14\\
Experiment \cite{piezo} & 3.61 & 2.40 & 0.28 & 1.59\\
Experiment \cite{piezo2} & 3.76 & 2.43 & 0.23 & 1.33\\[0.8ex]
\hline \\[-1.8ex]
\end{tabular}}
\end{center}
\label{tab4}
\end{table}

\subsection{Photoelasticity}

In CRYSTAL14, using script {\tt PHOTOELA}, we calculate photoelastic
constants $p_{uv}$ accordingly Eq. (\ref{pockel}). By default, the
electronic contribution is evaluated in the high-frequency limit but
it is far from a threshold of electronic excitation. Pockels tensor
at finite frequency can be computed with option {\tt DYNAMIC}.

\begin{figure}[h!]
\centerline{\psfig{file=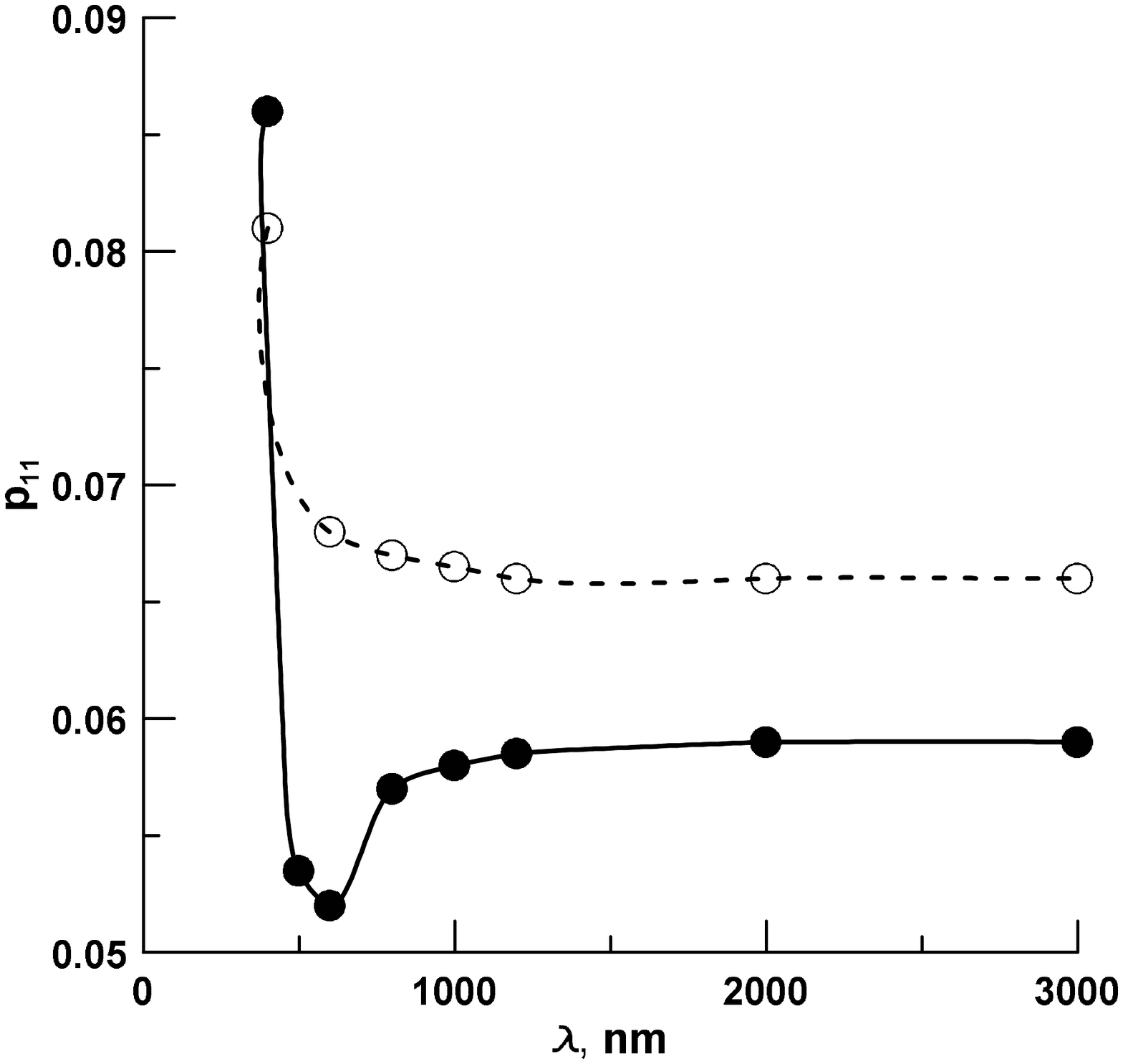,width=2.00in}\psfig{file=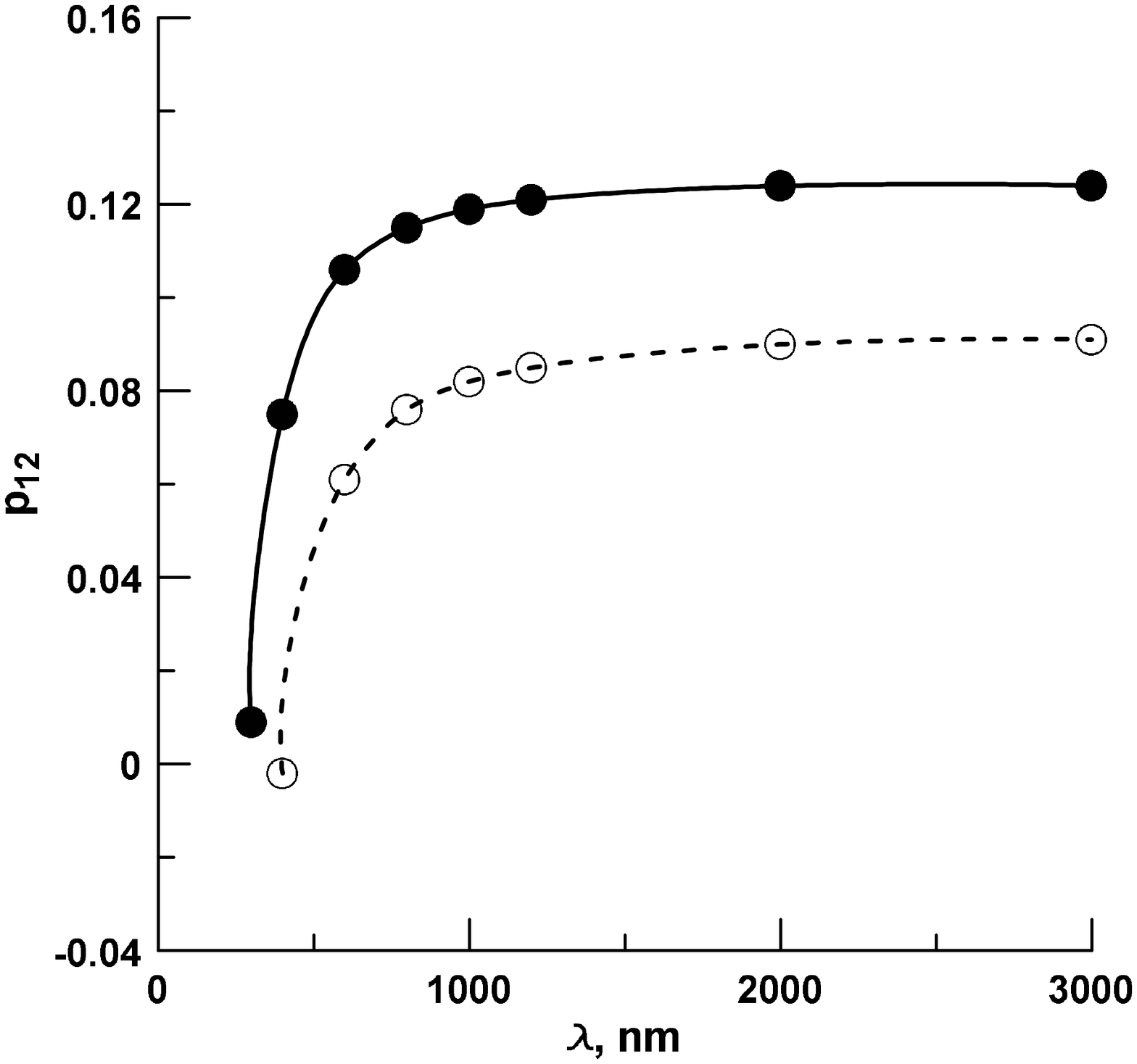,width=2.00in}}
\centerline{\psfig{file=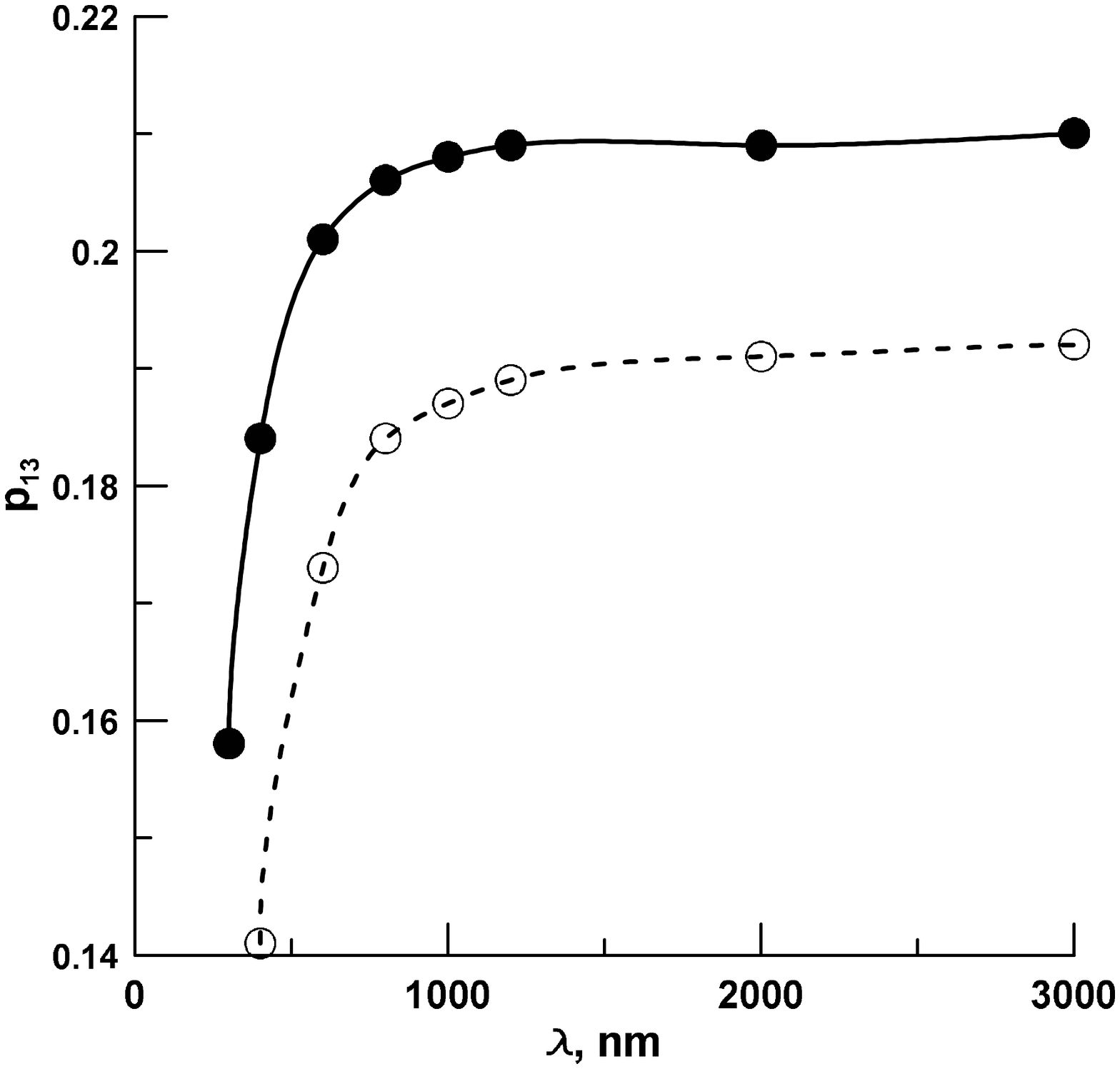,width=2.00in}\psfig{file=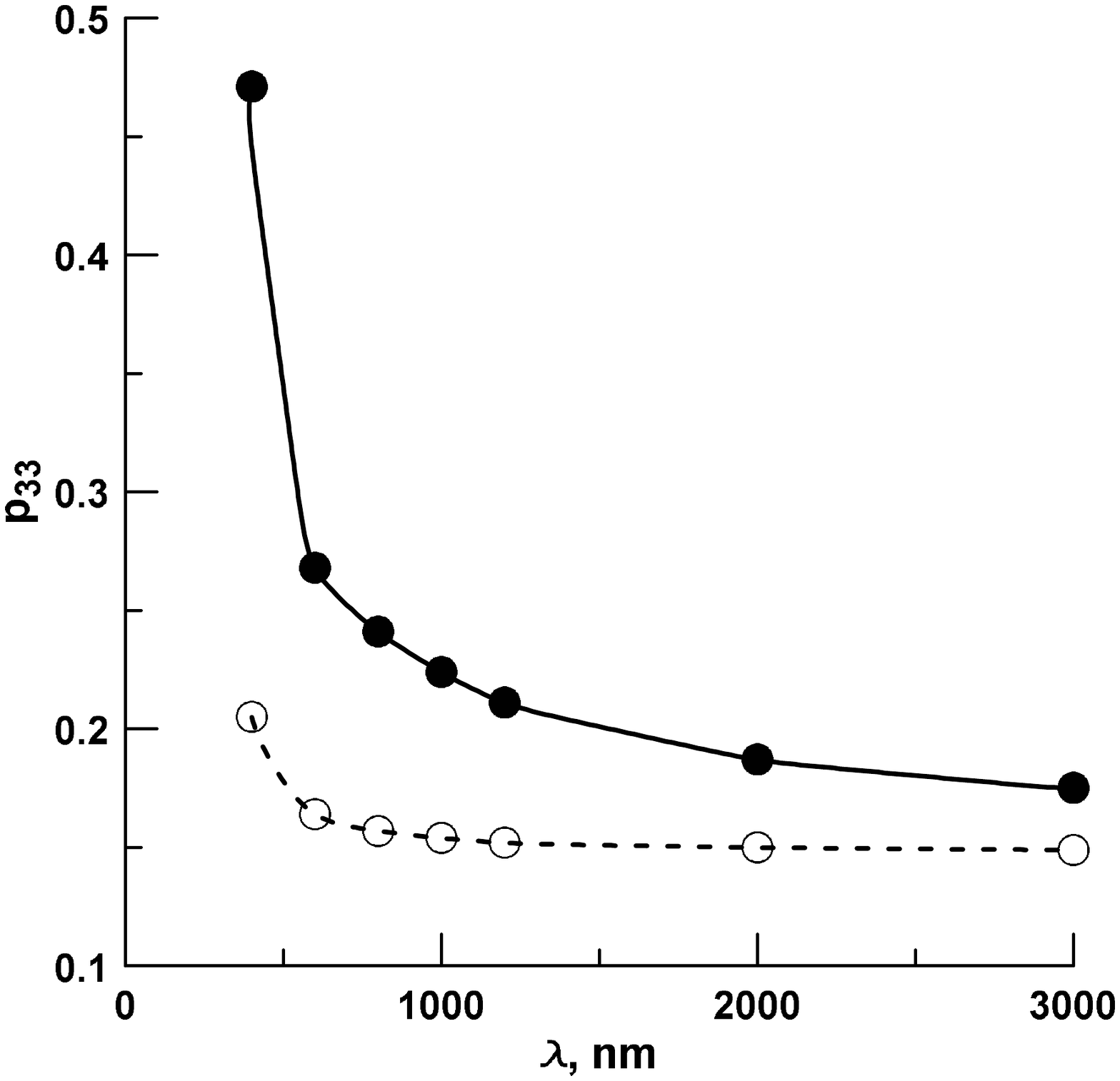,width=2.00in}}
\centerline{\psfig{file=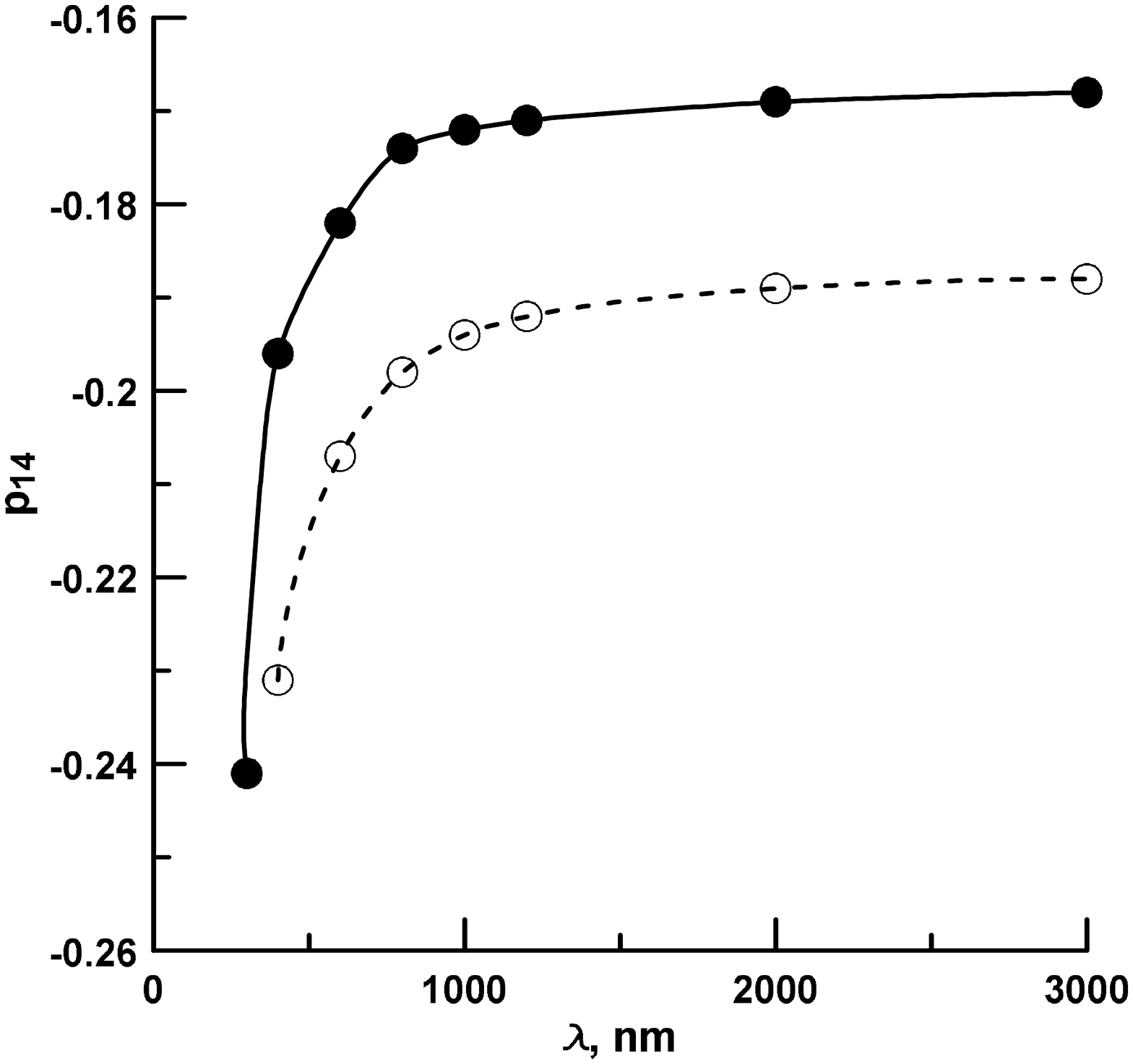,width=2.00in}\psfig{file=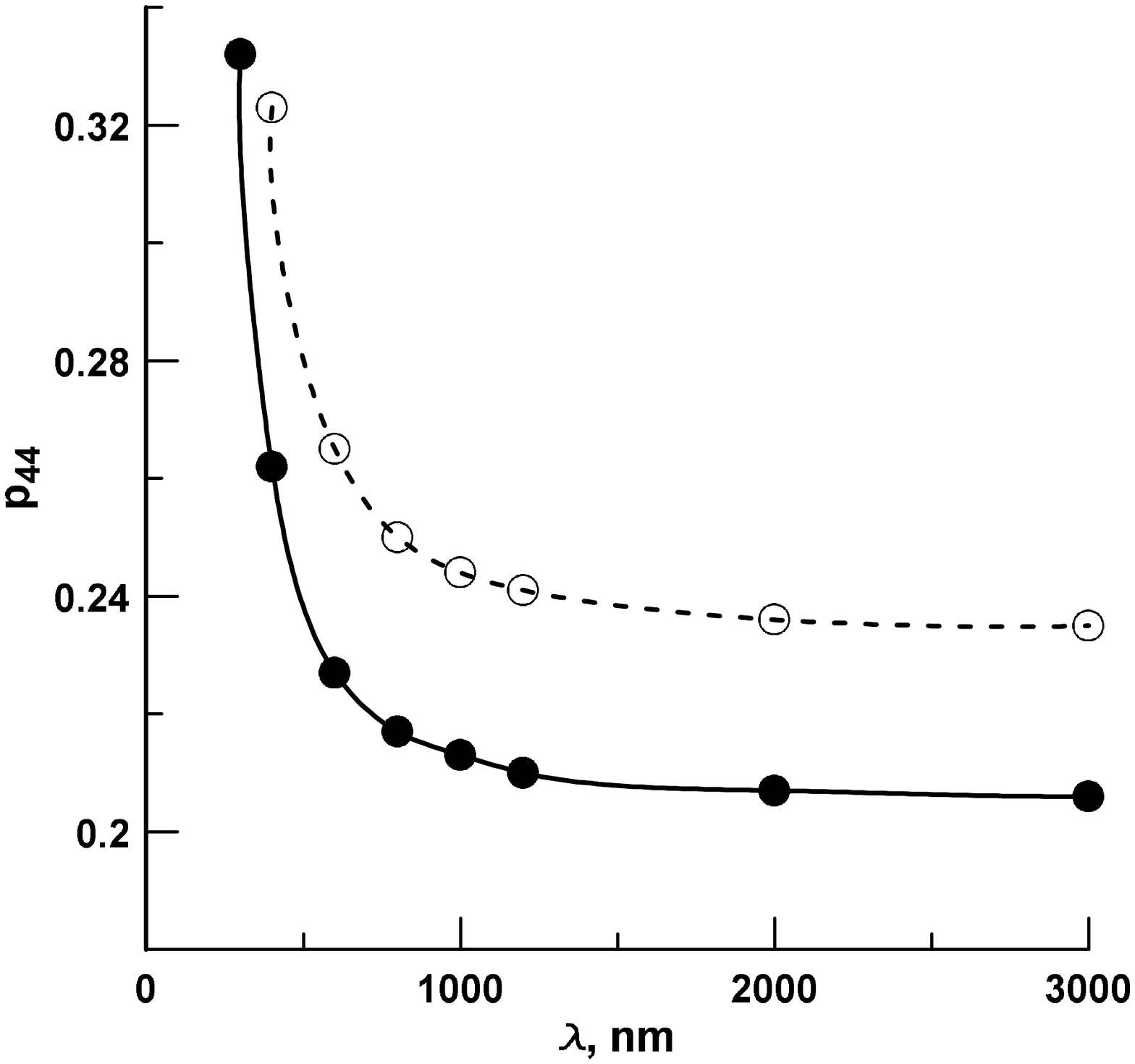,width=2.00in}}
\vspace*{8pt} \caption{The Pockels constants as function of
wavelength $\lambda$. Solid and dashed  lines are interpolations of
the theoretical predictions with PBESOL0 and PBESOL functionals.}
\label{fig:pockel}
\end{figure}

\begin{table}[h!]   
\caption{Photoelastic constants $p_{uv}$ in the high-frequency
limit.} \begin{center} {\begin{tabular}{@{}lcccccc@{}} \hline
\\[-1.8ex]
EC functional & $p_{11}$& $p_{12}$ & $p_{13}$ & $p_{33}$ & $p_{14}$ & $p_{44}$\\[0.8ex]
\hline \\[-1.8ex]
PBESOL & 0.067 & 0.092 & 0.192  & 0.161 & -0.187 & 0.234 \\
PBESOL0 & 0.059 & 0.124 & 0.209 & 0.175& -0.168 & 0.206\\
Experiment \cite{LN2009} & -0.021& 0.060 & 0.172 & 0.118 & -0.052 & 0.121\\
Experiment \cite{pockels} & 0.034& 0.072 & 0.139 & 0.060 & 0.066 & 0.300\\[0.8ex]
\hline \\[-1.8ex]
\end{tabular}}
\end{center}
\label{tab5}
\end{table}
{In practice, it is difficult or even impossible to determine
photoelastic constants directly, by measuring a strain induced
optical refraction, due to a problem of precise determination of a
crystal sample internal stress tensor.} The alternative
acousto-optical methods are ambiguous in the determination of the
sign of photoelastic constants. That is why the experimental data
strongly differ from each other. For instance, we obtain
theoretically that $p_{uv}=p_{vu}$, but in some experimental data
sets one has $p_{uv}\neq p_{vu}$. The full collection of
experimental data for photoelastic constants is presented in Ref.
\cite{LN2009}. {The additional problem to compare with the data is a
frequency-dependence of Pockels constants}. In Fig. \ref{fig:pockel}
we plot our predictions obtained with hybrid PBESOL0 functional.
{Throughout} the interval $300< \lambda< 1000$ nm, the photoelastic
constants change rapidly, and only at the $\lambda>1000$ nm it is
possible to neglect their dependence from wavelength. We should note
that our calculations can be uncorrect at the wavelength smaller 300
nm and it may be a reason of irregular behavior of $p_{11}$ in this
region. The estimated values of optical band gap for LN crystal with
different functionals are $E_{gap}=5.11$ eV (PBESOL0) and
$E_{gap}=3.10$ eV (PBESOL). The wavelength $\lambda=300$ nm
corresponds photon energy $h \nu=4.13$ eV, which is very close to
band gap energy, {so} the absorption effect can not be neglected.

\subsection{Refraction in high-frequency limit}

Dielectric tensor in high-frequency limit is calculated in CRYSTAL14
program automatically using script {\tt CPKS}. Let us note what we
mean here by "high-frequency limit". It is a region of frequencies
where ionic contribution is negligibly small, {it allows to} neglect
a slow response of heavy ions instead of fast electronic component.
From the other side, considered frequencies should be smaller than
ones needed for electronic excitation in separate atoms. Using
script {\tt PHOTOELA} with option {\tt DYNAMIC} we can calculate
dielectric tensor at finite frequency from this "high-frequency"
region.

\begin{figure}[h!]
\centerline{\psfig{file=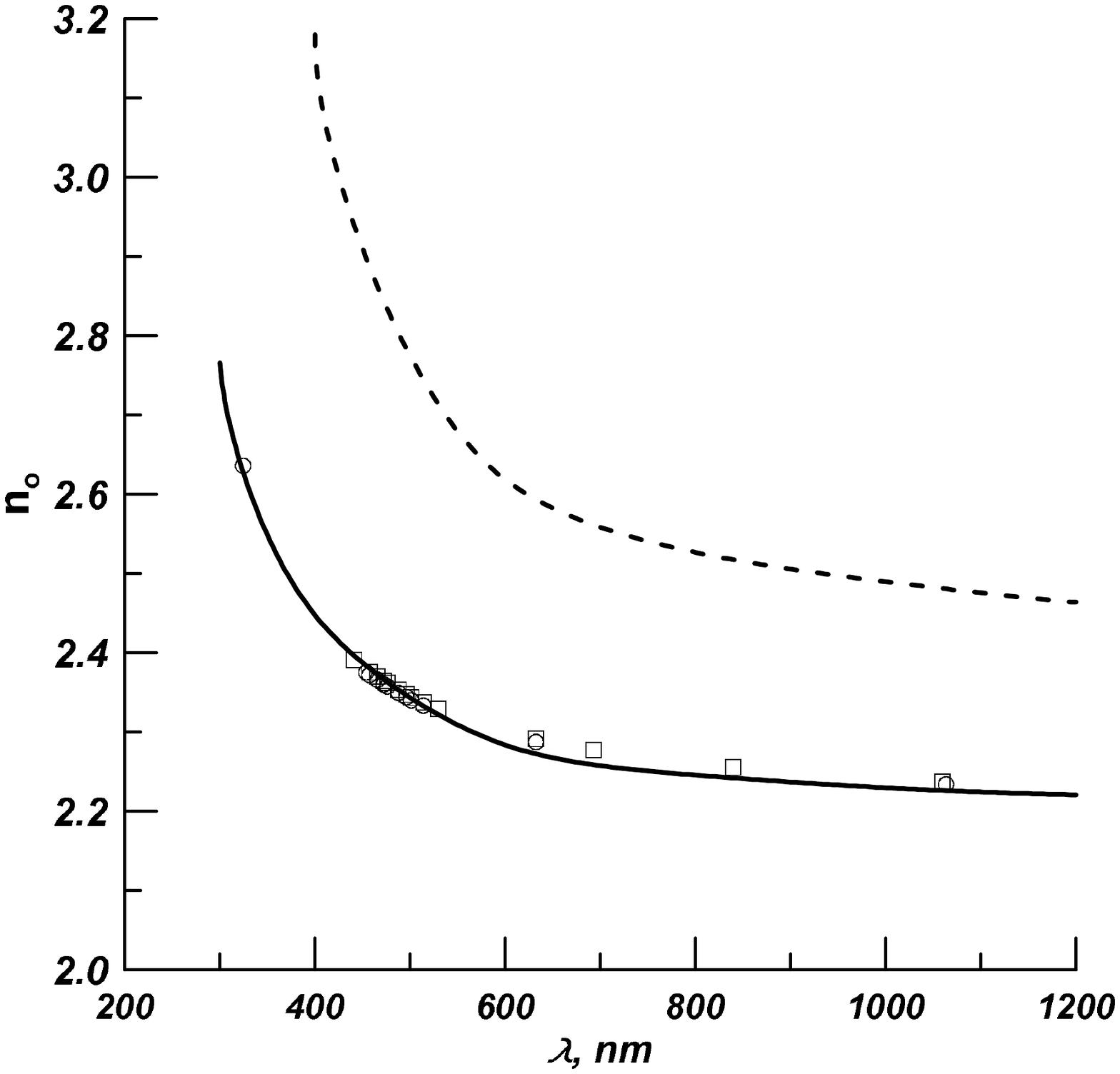,width=3.00
in}\psfig{file=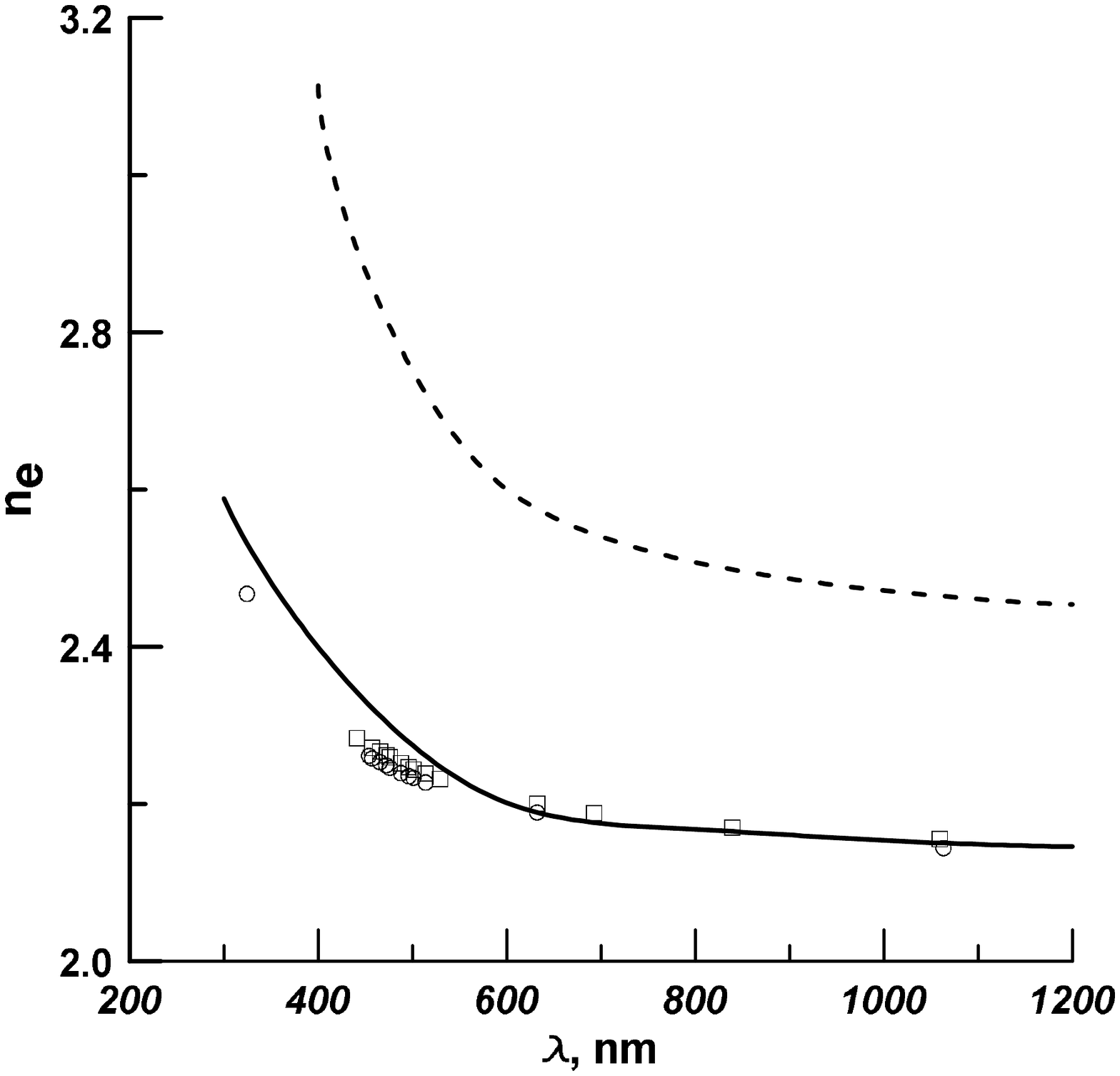,width=3.00 in}}
\vspace*{8pt} \caption{The ordinary $n_o$ and extraordinary $n_e$
refractive indexes as function of wavelength $\lambda$. Solid and
dashed lines are the theoretical predictions with PBESOL0 and PBESOL
functionals, points are experimental data from
Refs.~\protect\cite{refrac,refrac2}} \label{fig:refrac}
\end{figure}

Our calculation with PBESOL0 functional reproduces well the measured
precise dependence from wavelength for ordinary ($n_o=n_{zz}$) and
extraordinary ($n_e=n_{xx}=n_{yy}$) refractive indices in principal
axes system, as it is shown in Fig. \ref{fig:refrac}. For the
accurate comparison, we collect calculated in the high-frequency
limit refractive indices and experimental data for $\lambda=1150$ nm
in the Table \ref{tab3}.The absorption of light in this region is
negligibly small.

\begin{table}[h!]   
\caption{Refractive indices $n_o$ and $n_e$ in high-frequency
limit.} \begin{center} {\begin{tabular}{@{}lcc@{}} \hline
\\[-1.8ex]
EC functional & $n_{e}$& $n_{o}$\\[0.8ex]
\hline \\[-1.8ex]
PBESOL & 2.415&2.433 \\
PBESOL0 & 2.161&2.214\\
Experiment \cite{refrac} ($\lambda=1150$ nm)& 2.150&2.232\\[0.8ex]
\hline \\[-1.8ex]
\end{tabular}}
\end{center}
\label{tab3}
\end{table}

\subsection{Infrared and far-infrared regions}

\begin{figure}[h!]
\centerline{\psfig{file=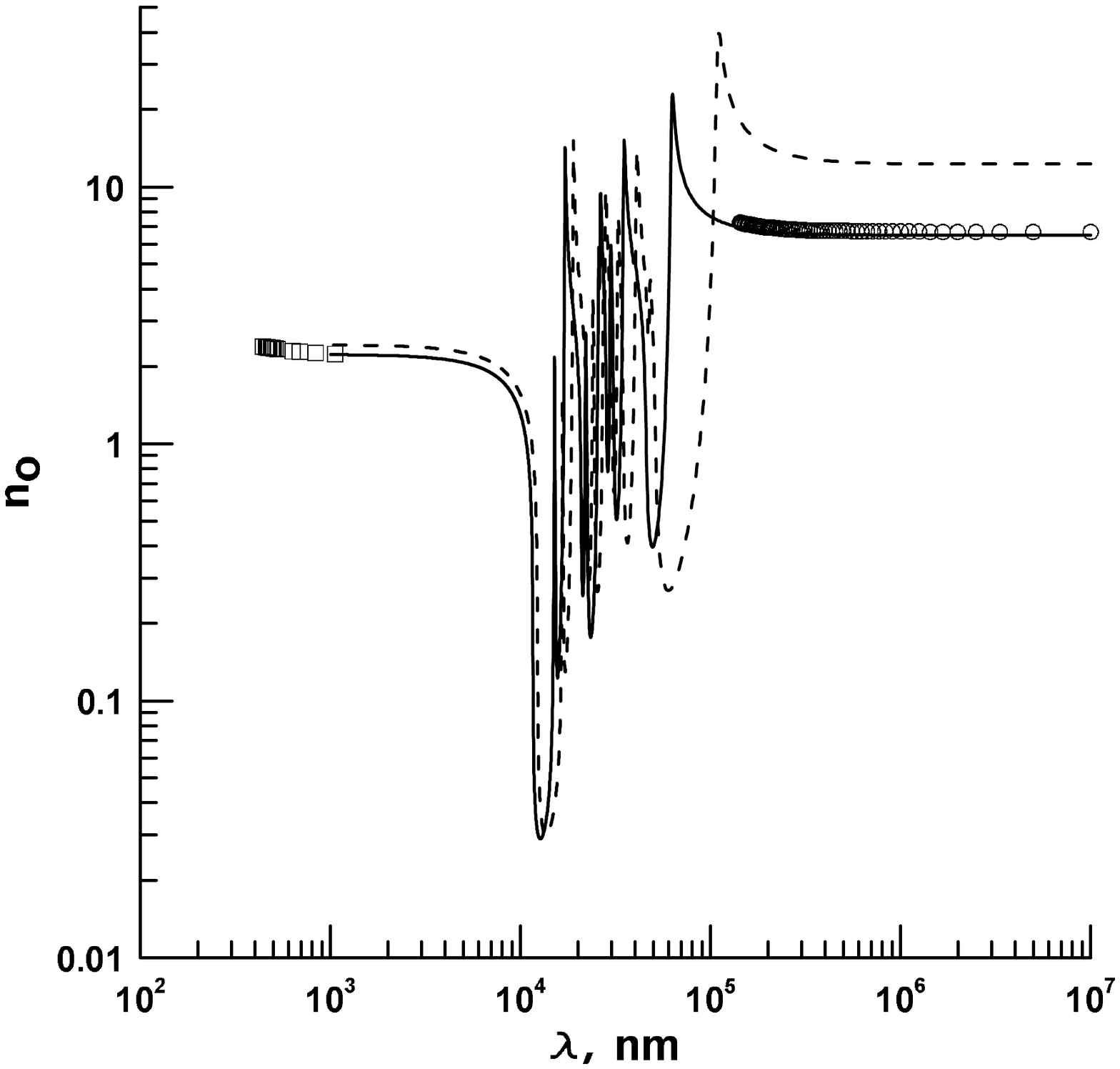,width=3.0
in}\psfig{file=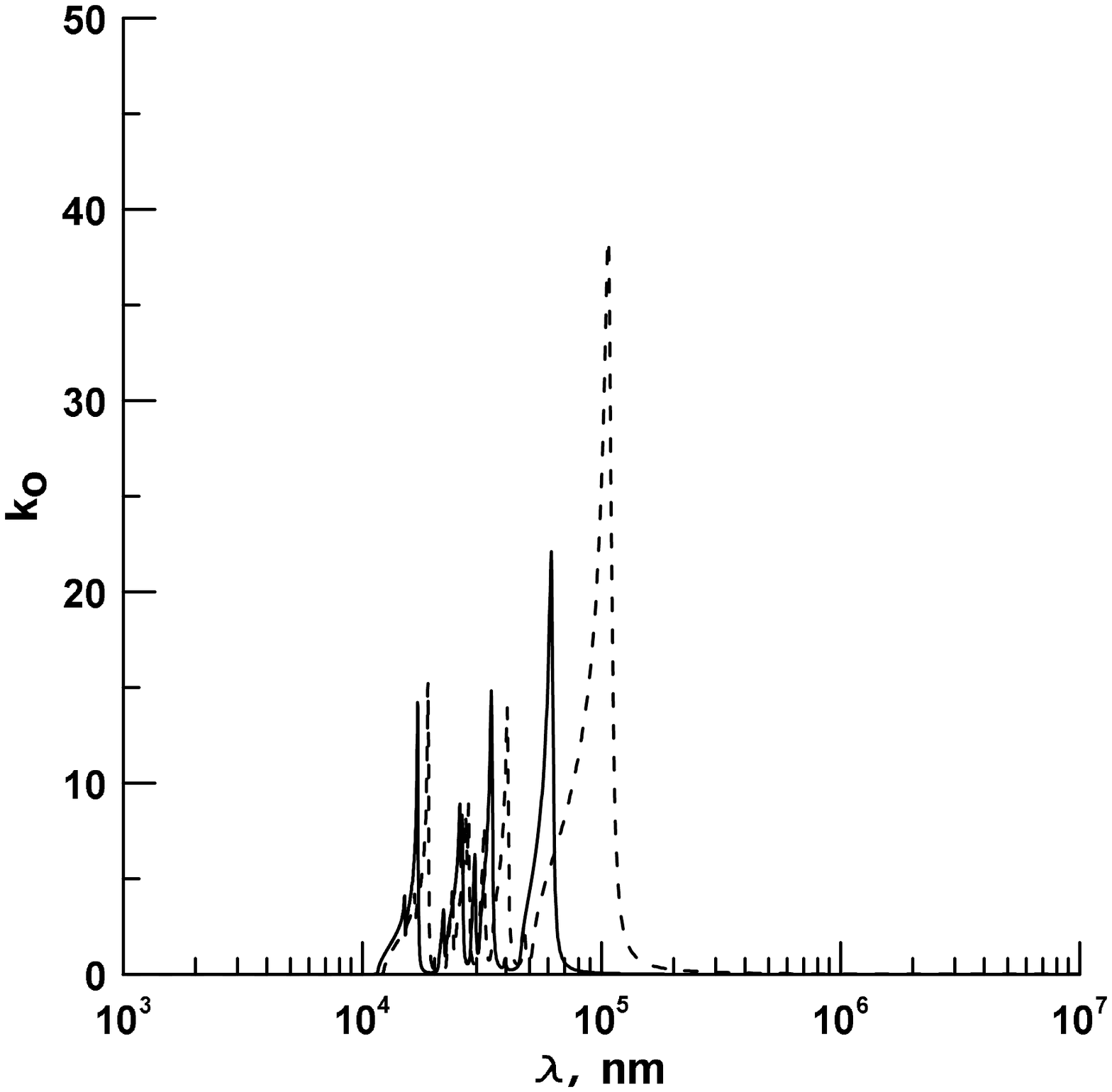,width=3.0 in}} \vspace*{8pt}
\caption{The real  and imaginary parts of refraction index ($n_o$
and $k_o$) as function of wavelength $\lambda$, obtained with PBESOL
(dashed line) and PBESOL0 (solid line) functionals. Experimental
data are from Refs. \protect\cite{refrac} (visible region) and
\protect\cite{LNTHz} (far-infrared region)} \label{fig:nreal}
\end{figure}

The results of our calculations for real and imaginary parts of
ordinary refraction index $n^*_o=n_o+i k_o$ are shown as function of
wavelength in Fig. \ref{fig:nreal}. We obtain approximate agreement
between data and theoretical prediction with PBESOL0 hybrid
functional in far-infrared region ($\lambda > 2\cdot 10^5$ nm)
{while the} prediction with PBESOL functional strongly overestimates
experimental data. As we predict, the infrared region of light
absorption of LN crystal lies between $2 \cdot 10^4$ nm and $2\cdot
10^5$ nm. {It is yet experimentally unknown region and a careful
study of the optical properties of LN crystal in this domain should
be very important for understanding of the vibrational and elastic
properties.}

\begin{figure}[h!]
\centerline{\psfig{file=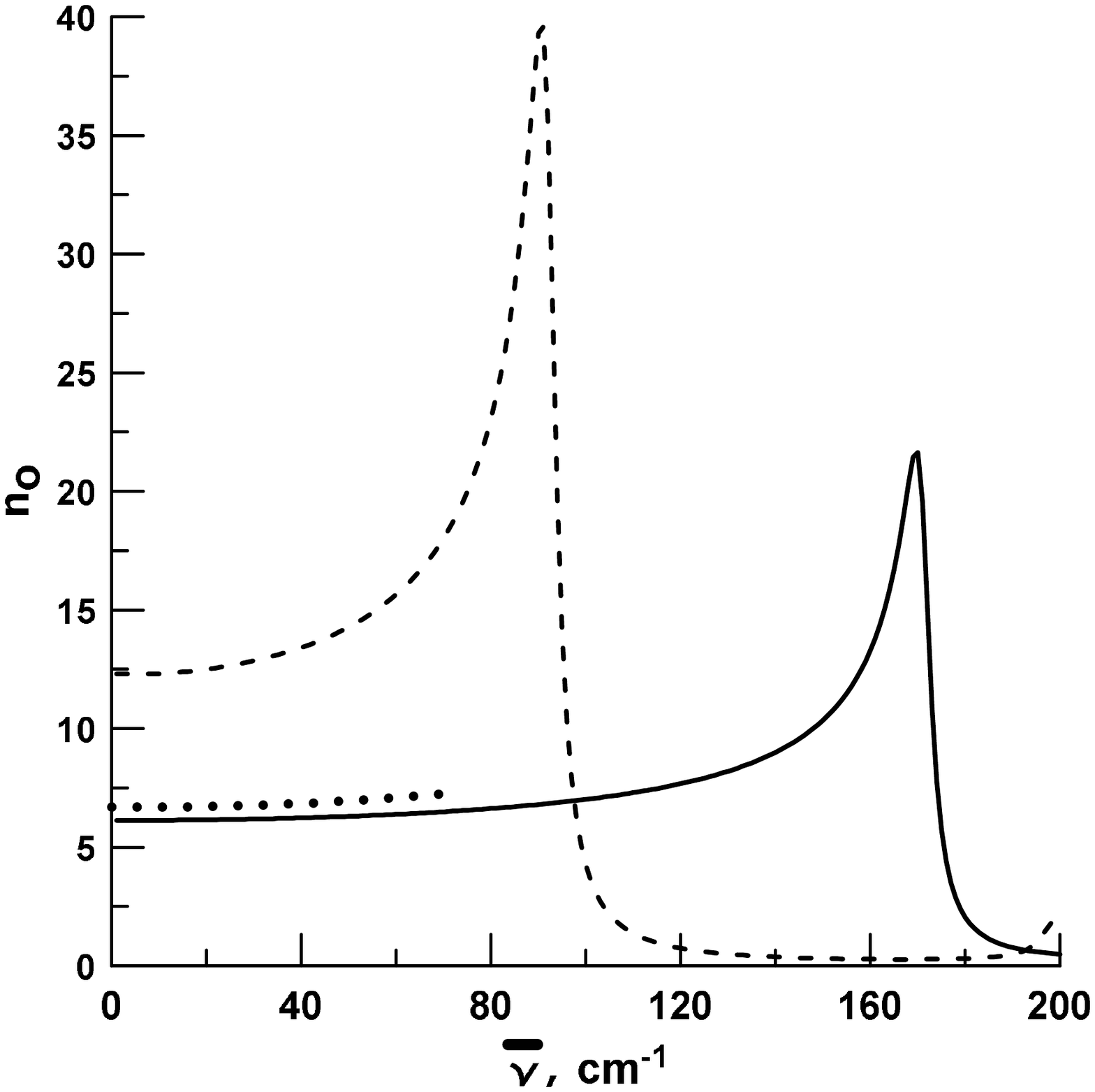,width=3.0
in}\psfig{file=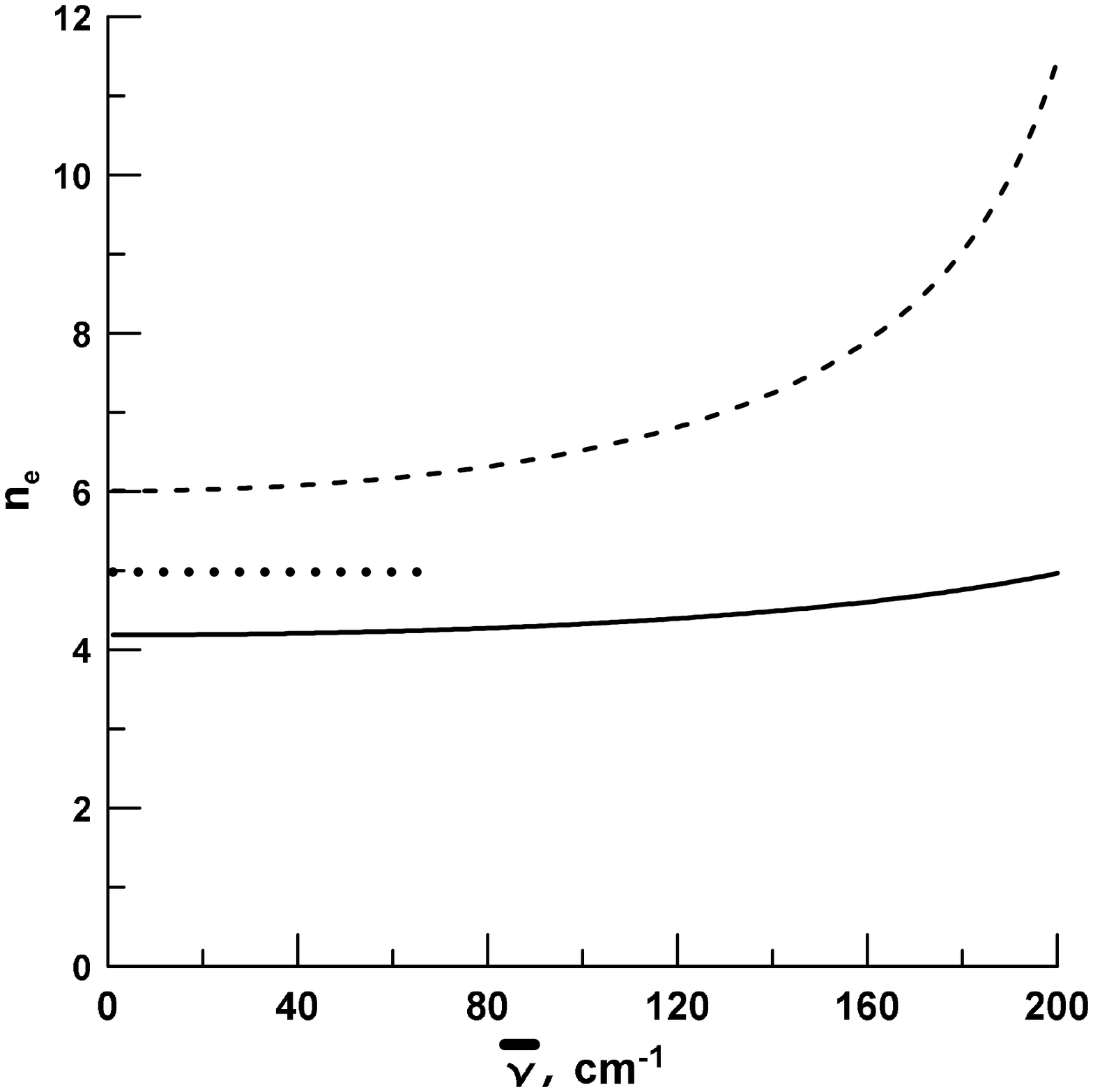,width=3.0 in}} \vspace*{8pt}
\caption{The refraction indices $n_o$ and $n_e$ as function of
wavenumber $\tilde{\nu}$, obtained with PBESOL (dashed line) and
PBESOL0 (solid line) functionals. Experimental data (dotes) are from
Ref. \protect\cite{LNTHz}} \label{fig:nosmall}
\end{figure}

For the precise comparison of our prediction for refractive indices
with existing experimental data we plot them as a function of
wavenumber in the region $2 \div 200$ cm$^{-1}$, see Fig.
\ref{fig:nosmall}. The approximation of experimental data
\cite{LNTHz} are plotted as doted lines. In case of extraordinary
polarization both functionals overestimate data at 20\% and 50 \%,
correspondingly. For ordinary polarization, hybrid functional
PBESOL0 predicts $n_o$ very close to data, with accuracy of {several
percents}.

\begin{figure}[h!]
\centerline{\psfig{file=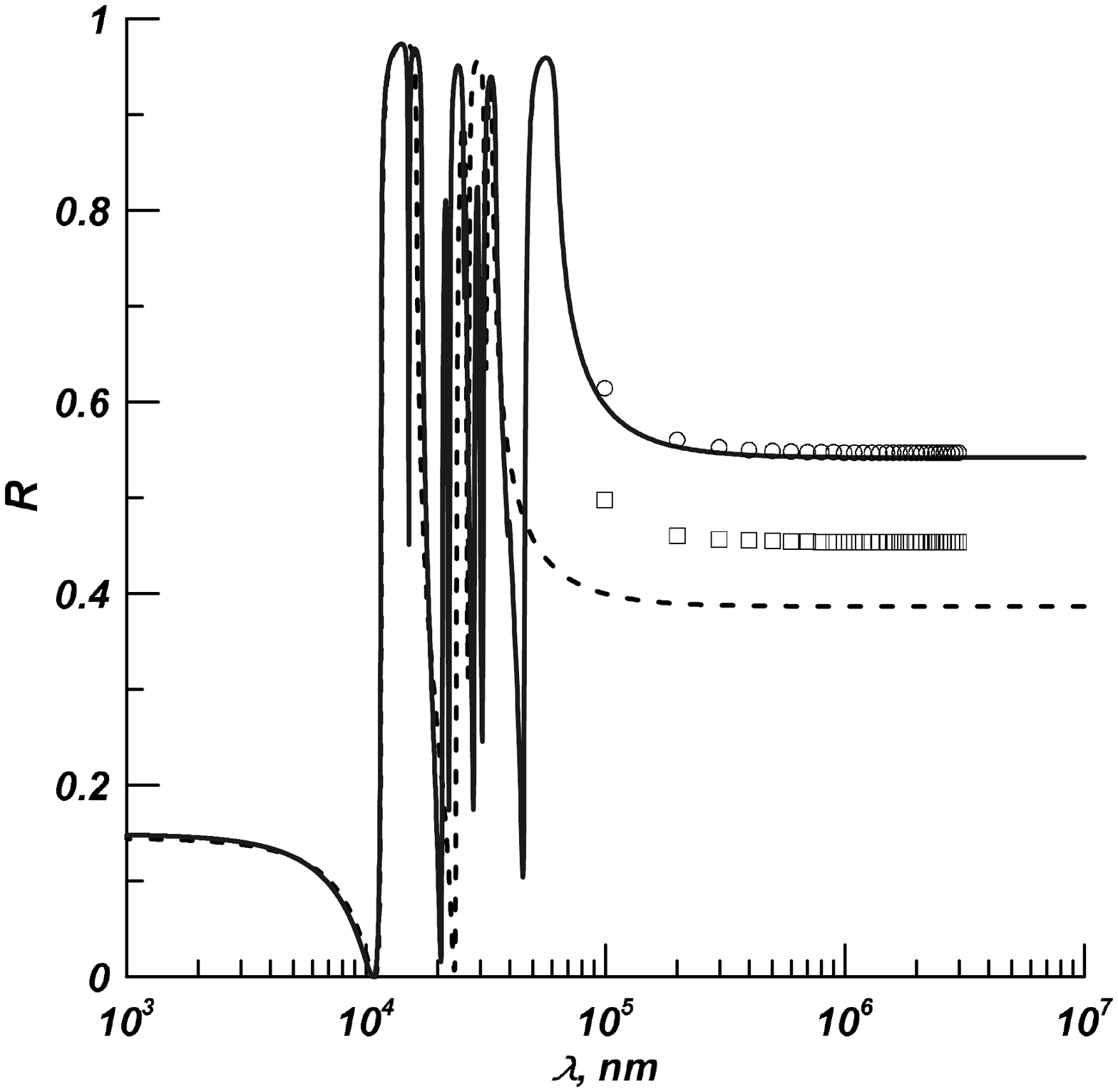,width=3.0 in}} \vspace*{8pt}
\caption{The reflection $R$ for ordinary and extraordinary
polarization as function of wavelength $\lambda$, obtained with
PBESOL0 functional, solid and dashed lines, correspondingly.
Experimental data are from Ref. \protect\cite{LNTHz}}
\label{fig:reflect}
\end{figure}

The reflection indexes $R$ are sufficiently different for ordinary
and extraordinary polarization in the far-infrared region. Our
calculation predicts even larger difference than it is measured
experimentally, as it shown in the Fig. \ref{fig:reflect}. The same
as for refractive indices we underestimate $R$ in case of
extraordinary polarization. It is interesting that at static limit
$\triangle R=R_0- R_e \simeq 0.1-0.2$ and, accordingly our
prediction with PBESOL0 functional, this difference {tends to be}
about $\triangle R\simeq 0.5-0.6$ in the region between $6\cdot
10^4$ nm and $10^5$ nm.


\section{Conclusions}

We demonstrate {self-consistency} in {\it ab initio} quantum
mechanical calculations of elastic, elasto-optical and optical
properties for ferroelectric LN crystal in the regions of low and
high frequencies and predict an extinction coefficient property of
LN crystal in the infrared region. The good agreement of DFT
calculations {with the use of} hybrid functional PBESOL0 with the
discussed here data is demonstrated.

\section*{Acknowledgements}

The work was funded by the Ministry of Education and Science of
Russia under Competitiveness Enhancement Program of Samara
University for 2013-2020, project 3.5093.2017/8.9.

\newpage

\begin{thebibliography}{0}

\bibitem{LN1985} R.S. Weis and T.K. Gaylord, {\it Appl. Phys.} {\bf A37}, 191 (1985).

\bibitem{LNwww} Electronic recourse: materials.springer.com

\bibitem{LN2002} M. Jazbinsek and M. Zgonik, {\it Appl. Phys.} {\bf
B74}, 407 (2002).

\bibitem{LN2009} A.S. Andrushchak et al., {\it J. Appl. Phys.} {\bf
106}, 073510 (2009).

\bibitem{LN65} A.S. Barker, Jr. and R. Loudon, {\it Physical
Review} {\bf 158}, 433 (1967).

\bibitem{LN68} I.P. Kaminov and D.W.Johnston Jr., {\it Phys. Rev.}
{\bf 160}, 519 (1967).

\bibitem{LN62} A. Ridah, M. D. Fontana and P. Bourson, {\it Phys. Rev.} {\bf B 56}, 5967 (1997).

\bibitem{LN63} A. Ridah, P. Bourson, M. D. Fontana and G. Malovichko, {\it J. Phys.: Condens.
Matter} {\bf 9}, 9687 (1997).

\bibitem{LN64} U. T. Schwarz and M. Maier, {\it Phys. Rev.} {\bf B 55}, 11041 (1997).

\bibitem{LN66} M. R. Chowdhury, G. E. Peckham and D. H. Saunderson, {\it J.Phys.C: Solid State
Phys.} {\bf 11}, 1671 (1978).

\bibitem{LN67} Y. Repelin, E. Husson, F. Bennani and C. Proust, {\it J. Phys. Chem. Solids} {\bf 60}, 819
(1999).

\bibitem{LN69} R. Claus, G. Borstel, E. Wiesendanger and L. Steffan, {\it Z. Naturforsch} {\bf A 27}, 1187
(1972).

\bibitem{LN70} X. Yang, G. Lan, B. Li and H. Wang, {\it Phys. Stat. Sol.} {\bf B142}, 287 (1987).

\bibitem{LNTHz} M. Schall, H. Helm and S.R. Keiding, {\it Int. J. Infrared and Millimeter Waves} {\bf 20},  595 (1999).

\bibitem{LNvisible} J. R. Schwesyg, M. C. C. Kajiyama, M. Falk, et
al., {\it Appl. Phys.} {\bf B100}, 109 (2010).

\bibitem{DFT1} P. Hohenberg, W. Kohn W. {\it Phys. Rev.} {\bf 136},
B864 (1964).

\bibitem{DFT2} W. Kohn, L.J. Sham, {\it Phys. Rev.} {\bf 140}, (1965).

\bibitem{PRB2000} K. Parlinski, Z.Q. Li and Y. Kawazoe, {\it Phys. Rev.} {\bf B 61}, 272(2000).

\bibitem{PRL2004} M. Veithen, X. Gonze, and P. Ghosez, {\it Phys.
Rev. Lett.} {\bf 93}, 187401 (2004).

\bibitem{PRB2008} W. G. Schmidt, M. Albrecht, S. Wippermann et al.,
{\it Phys. Rev.} {\bf B77}, 035106 (2008).

\bibitem{CMS2013} S. Mamoun, A.E. Merad and L. Guilbert, {\it Comp. Mat.
Sci.} {\bf 79}, 125 (2013).

\bibitem{PRB2014}  Ya. Li, W.G. Schmidt, and S. Sanna, {\it Phys.
Rev.} {\bf B89}, 094111 (2014).

\bibitem{China2015} S. Dan-Dan, W. Qing-Lin, H. Chong, C. Kai, and
P. Yue-Wu, {\it Chin. Phys. } {\bf B24}, 077104 (2015).

\bibitem{LNGW2016} A. Riefer, M. Friedrich, S. Sanna, U. Gerstmann, A. Schindlmayr, and W. G. Schmidt,
{\it Phys. Rev.} {\bf B93}, 075205 (2016).

\bibitem{cry14} R. Dovesi et al., {\it Int. J. Quantum. Chem.} {\bf 114}, 1287 (2014).

\bibitem{cry14_2} F. Pascale, C. M. Zicovich-Wilson, F. Lopez, B. Civalleri,
R. Orlando, R. Dovesi, {\it J. Comp. Chem.} {\bf 25}, 888 ( 2004).


\bibitem{cry14_3} M. Ferrero, M. Rerat, R. Orlando, R. Dovesi, {\it J. Comp. Chem.} {\bf 29}, 1450 (2008).

\bibitem{Erba2013} A. Erba, R. Dovesi, {\it Phys. Rev.} {\bf B88},
045121 (2013).

\bibitem{elast} A. Erba, A. Mahmout, R. Orlando, R. Dovesi, {\it Phys. Chem. Minerals.} {\bf 41}, 151 (2014).

\bibitem{piezot} A. Erba, Kh. E. El-Kelany, M. Ferrero, I. Baraille, and M. Rerat. {\it Phys. Rev.} {\bf B88}, 035102
(2013).

\bibitem{pbesol} J. P. Perdew, A. Ruzsinszky, G. I. Csonka, O. A. Vydrov, G. E. Scuseria,
L. A. Constantin, X. Zhou, and K. Burke. {\it Phys. Rev. Lett.} {\bf
100}, 136406 (2008).



\bibitem{baranek} G. Sophia, P. Baranek, C. Sarrazin, M. Rerat, R. Dovesi,
  {\it Phase Transitions: A Multinational Journal} {\bf 81}, 1069
  (2013).

\bibitem{latice} A.V. Postnikov, V. Caciuc, G. Borstel, {\it J. Phys. and Chem.
Sol.} {\bf 61}, 295 (2000).



\bibitem{cab} A.W. Warner, M. Onoe, and G.A. Coquin, {\it J. Acoust. Soc.
Am} {\bf 42}, 1223 (1967).

\bibitem{refrac}  G.D. Boyd, W.L. Bond, H.L. Carter, {\it J. Appl.
Phys.} {\bf 38}, 1941 (1967).

\bibitem{piezo} I.A. Dankov, E.F. Tokarev, G.S. Kudryashov, and K.G.
Belobaev, {\it Inorg. Mater.} {\bf 19}, 1049 (1983).

\bibitem{piezo2} R.T. Smith, and F.S. Welsh, {\it J. Appl. Phys.} {\bf 42}, 2219 (1971).

\bibitem{pockels} R.J. O'Brein, G.J. Rosasco and A. Weber, {\it J.
Opt. Soc. Am.} {\bf 60}, 14 (1970).

\bibitem{refrac2} D.H. Junt, M.M. Fejer, and R.L. Byer, {\it J.
Quant. Elec.} {\bf 26}, 135 (1990).

\end{thebibliography}
\end{document}